\documentclass[conference]{IEEEtran}
\usepackage{color}

\usepackage{epsf}
\usepackage{times}
\usepackage{epsfig}
\usepackage{epstopdf}
\usepackage{graphicx}
\usepackage{algorithmic}
\usepackage{amsmath}
\usepackage{amssymb}
\usepackage{amsxtra}
\usepackage{multirow}
\usepackage{mathtools}
\usepackage{algorithm}
\usepackage{subcaption}
\usepackage{multirow}
\usepackage{comment}
\usepackage[normalem]{ulem}
\usepackage{color}
\usepackage{epsf}
\usepackage{times}
\usepackage{epsfig}
\usepackage{epstopdf}
\usepackage{graphicx}
\usepackage{algorithm}
\usepackage{algorithmic}
\usepackage{amsmath}
\usepackage{amssymb}
\usepackage{amsxtra}
\usepackage{multirow}
\usepackage{mathtools}
\usepackage{subcaption}
\usepackage{multirow}
\usepackage{comment}
\usepackage[normalem]{ulem}
\usepackage{tabu}
\usepackage[dvipsnames]{xcolor}
\usepackage{multicol}
\usepackage{cite}
\usepackage{caption}
\newcommand\Mycomb[2][^n]{\prescript{#1\mkern-0.5mu}{}C_{#2}}

\begin{document}

\title{A Stochastic Team Formation Approach for Collaborative Mobile Crowdsourcing}

\author{\IEEEauthorblockN{Aymen Hamrouni$^1$, Hakim Ghazzai$^1$, Turki Alelyani$^2$, and Yehia Massoud$^1$}
\IEEEauthorblockA{\small $^1$School of Systems  \& Enterprises, Stevens Institute of Technology, Hoboken, NJ, USA\\
Email: \{ahamroun, hghazzai, ymassoud\}@stevens.edu} 
$^2$College of Computer Science and Information Systems, Najran University, Najran, Saudi Arabia \\
Email: tnalelyani@nu.edu.sa
\vspace{-0.2cm}}

\maketitle
\thispagestyle{empty}

\begin{abstract}
\boldmath{
Mobile Crowdsourcing (MCS) is the generalized act of outsourcing sensing tasks, traditionally performed by employees or contractors, to a large group of smart-phone users by means of an open call. With the increasing complexity of the crowdsourcing applications, requesters find it essential to harness the power of collaboration among the workers by forming teams of skilled workers satisfying their complex tasks' requirements. This type of MCS is called Collaborative MCS (CMCS). Previous CMCS approaches have mainly focused only on the aspect of team skills maximization. Other team formation studies on social networks (SNs) have only focused on social relationship maximization. In this paper, we present a hybrid approach where requesters are able to hire a team that, not only has the required expertise, but also is socially connected and can accomplish tasks collaboratively. Because team formation in CMCS is proven to be NP-hard, we develop a stochastic algorithm that exploit workers knowledge about their SN neighbors and asks a designated leader to recruit a suitable team. The proposed algorithm is inspired from the optimal stopping strategies and uses the odds-algorithm to compute its output. Experimental results show that, compared to the benchmark exponential optimal solution, the proposed approach reduces computation time and produces reasonable performance results.}
\end{abstract}


\section{Introduction}
\label{Sec1a}
Mobile Crowdsourcing (MCS) utilizes the power of mobile devices to accomplish specific sensing and data collection tasks without requiring pre-deployed dedicated infrastructure. Typically, MCS is composed of three parties: task requesters, task workers, and a cloud platform. When a task requester finds difficulties in collecting certain information, he/she can initiate a crowdsourcing task describing his/her problems and then, announce it via the platform to the crowd. The platform will be in charge in selecting, according to certain criteria, the group of appropriate contributors that can deliver satisfying results. Existing MCS approaches include simple tasks that require selected workers to complete what is necessary independently of each other (e.g. traffic monitoring~\cite{8843858}). Due to the diversity of workers’ skills on performing tasks, many researchers, for example~\cite{7446292}, mainly focus on whether the hired workers are professional enough such that they can satisfy the task's skill requirements.

In many mobile crowdsourcing applications, the class of tasks in question, also called projects, can be so complex that the success of their completion depends on not only the expertise of the hired workers but also on how efficiently these workers can work together as a team. This could be, for example, the case of a storm emergency evacuation situation in which a group of people is supposed to provide up-to-the minute information about shelters and evacuation routes. If the communication fails between the workers for one reason or another (e.g., language barriers or geographic distance), the job cannot be achieved on time. Therefore, besides having the required skills, the success of the project depends on how efficiently the team members are able to communicate. To combine and fulfill these needs, a suitable team recruitment process must be put in place that: (i) recruits workers with a set of skills required by the project and (ii) ensures that they can effectively collaborate, communicate and work together as a team.

MCS is a very useful paradigm to help requesters access the power of human resources and mobile devices to complete projects that are difficult for computers~\cite{article5,article6}. In traditional crowdsourcing applications, workers are recruited and asked to complete the same task independently of each other and without any contact, e.g., covering an ongoing event by taking pictures and uploading them to a MCS framework for handover~\cite{1111}, or improving the labeling accuracy and completing as many labeling tasks as possible in web-based crowdsourcing platforms~\cite{7875084}. However, with the increasing complexity of some tasks, recent studies have begun to address the need to consider recruiting a team of workers~\cite{7248382,AAAI1612106}. In fact, some approaches, such as~\cite{116}, focused on dividing complex tasks into flows of simple sub-tasks and allocating these sub-tasks to a team of workers. At the end, the partial results are combined to produce the overall outcome. These approaches focus only on the expertise of recruited team and does not consider the interaction within members. Other approaches focused on team formation in social networks and proposed a solution to hire teams with good social relationships indifferent of the members' level of expertise~\cite{article8}.

To complement these studies, we aim, in this paper, to present a hybrid crowdsourcing recruitment approach where the cloud platform selects a leader to which it delegates the team formation procedure. The objective is to form not only a skilled but also socially connected team. The proposed approach overcomes the limited knowledge of the platform about the workers' skills and profiles. It also relies on leaders who usually have a better knowledge about the workers in their SN neighborhood. To form the team, a probabilistic recruitment algorithm that considers the team members parameters (e.g. degree of expertise, social relationship, recruitment confidence level, and financial cost) from the leader point of view. Selected simulation results show that our proposed probabilistic algorithm reduces computation time and produces close performances to the optimal benchmarking algorithm.

\section{MCS Model}\label{Sec2}

\begin{figure}[t]
    \centering\vspace{0.2cm}
    \includegraphics[width=9cm]{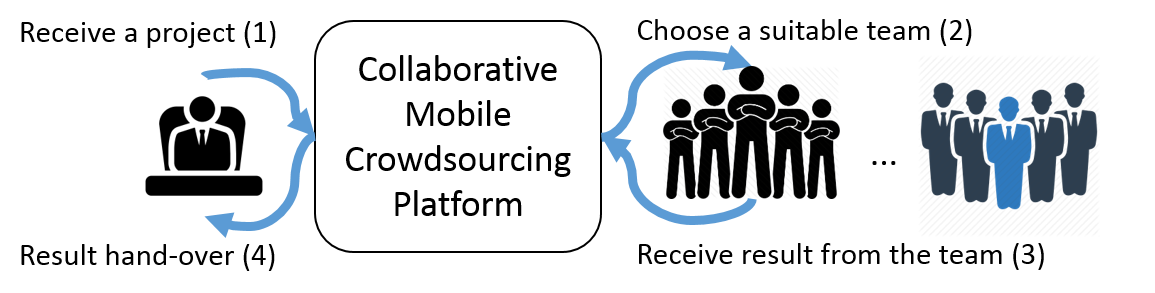}
    \caption{Work-flow of a CMCS platform.}
    \label{cmcsplatform}\vspace{-0.0cm}

\end{figure}

A CMCS system is composed of two external parties in addition to the cloud platform: the project initiator and the workers as shown in Fig.~\ref{cmcsplatform}. When a project initiator needs services, he/she submits its MCS project, having $\mathcal S_p$ as a set of required skills, to the platform. The latter is responsible of recruiting a suitable team that is capable of completing the project given the requirements of the former. After completing the project, the team submits its response to the platform for eventual hand-over to the project initiator.

We denote by $\mathcal W$ the set of $N$ workers registered in the CMCS platform where $\mathcal W=\{w_1, \dots, w_N\}$. Let $\mathcal S=\{s_1,\dots, s_M\}$ be the set of $M$ all possible skills that characterize workers. Each worker $w_i \in \mathcal W$ has a degree of expertise in skill $s_j \in \mathcal S$ denoted by $S_{ij}$ where $0\leq S_{ij} \leq 1$. The value of $S_{ij}$ can be interpreted as follows: $S_{ij}\rightarrow 1$ means that the worker $w_i$ is an expert in skill $s_j$. Otherwise, $S_{ij} \rightarrow 0$ means that worker $w_i$ does not have sufficient knowledge about skill $s_j$. Let $\mathcal S_i=\{S_{i1}, \dots, S_{iM}\}$ be the set of skills value provided by worker $w_i$. To execute a task with skill $j$, a worker $w_i$ may request a certain cost denoted by $C_{w_i,j}$.

We assume that the workers in the platform are modeled as an undirected and weighted graph $\mathcal G(\mathcal W,\mathcal E)$. Every node of $\mathcal G$ corresponds to a worker $w_i \in \mathcal W$ while the set of edges $ \mathcal E$ represents the SN relationships between the workers. Initially, we only consider the edges connecting a pairwise of workers that can directly communicate and collaborate and we associate to their weights the value $1$. Then, the edges between the remaining pairwise of nodes, e.g., $(w_i,w_j)$, which are not directly connected are given a  weight computed using the shortest number of hops, denoted by $n^{hops}_{w_i,w_j}$, needed for one of the pairwise nodes to reach the other. Hence, the graph $\mathcal G$ is converted into a mesh graph where all nodes are connected and the values of the edges' weights indicate the social relationship levels between each pair of workers. The values on each edge between two workers $w_i$ and $w_j$ is given as: $R_{w_i,w_j}=\frac{1}{1+n^{hops}_{w_i,w_j}}$. If an isolated sub-graph exists, then the weights connecting a node of this sub-graph to other external nodes is set to zero ($n_{hops}\rightarrow \infty$).

\section{Collaborative Team Formation Framework}
\label{sec3}
In this section, we present the collaborative framework that forms a team addressing the requirement of the outsourced project from the available workers. The platform aims to recruit a team based on the knowledge of its leader, i.e. given its social friendship, knowledge, and confidence level in recruitment. In other words, the platform needs to choose the leader and his/her team at the same time.

In order to complete the outsourced project defined $\mathcal S_p$, a chosen team $\mathcal T_L \in \Theta_L$ associated to a leader $L \in \mathcal W $ is defined as a set of workers that must cover all these skills where $\Theta_L$ is the set of all possible team combinations having as leader $L$. The cardinality of $\Theta_L$ is equal to $\Mycomb[|\mathcal W|-1]{|\mathcal S_p|-1}$ where $\Mycomb[m]{k}$ is the combination of selecting $k$ items from $m$ items. We suppose that each skill in $\mathcal S_p$ needs to be covered by exactly one worker $\in \mathcal T_L$, and each worker can provides only one skill. Using this definition, a possible team for the project is composed of any distinct random workers and their leader $L$. Hence, $|\mathcal T_L|=|\mathcal S_p|$.  

We denote by $\mathcal S_L(\mathcal T_L)$ the set containing all the possible combination of skills of the team $\mathcal T_L$ where $|\mathcal S_L(\mathcal T_L)|=(|S_p|\,!)$. Hence, the objective of the problem is to determine the leader $L \in \mathcal W$, his/her team $\mathcal T_L \in \Theta_L$ and the skill combination  $s_L(\mathcal T_L) \in\mathcal S_L(\mathcal T_L)$ in maximizing the following team efficiency metric denoted by $\text{TE}$:
\begin{align}
\text{TE}(L,\mathcal T_L,s_L(\mathcal T_L))= & \sum_{w \in  T_L} \hspace{-0.2cm} \bigg( \gamma_1 \hat{S}^L_{w,\bar{j}} - \gamma_2 U^{L}_w-\gamma_3 C_{w,\bar{j}}  \bigg) \notag \\
&+ \gamma_4 \sum_{\substack{(w_1,w_2) \in \mathcal T_L \times \mathcal T_L \atop w_1\neq w_2}}\hspace{-0.3cm}
R_{w_1,w_2},
\label{scoreequation}
\end{align}

This team efficiency expression includes four terms:\\
1) The first term that the platform aims to maximize is $\sum_{w \in  T_L}\hat{S}^L_{w,\bar{j}}$. It measures the total skill level of the team if the combination $s_L(\mathcal T_L)$ is chosen according to the knowledge of the leader $L$. In other words, we assume that the leader does not perfectly know the skill of each worker in the platform instead he/she knows an estimated value expressed as follows: $\hat{S}^L_{w,\bar{j}}=S_{w,\bar{j}}+\tilde{S}^L_{w}$ where $\tilde{S}^L_{w}$ is the error made by the leader given his knowledge about the worker $w$. This error can be modeled as a zero-mean distribution with a variance $U^L_w$. It reflects the confidence level of the leader when recruiting a worker. For example, its value decreases with the number of hops separating the leader and the worker in graph $\mathcal G$.\\
2) The platform aims also to minimize the second term $\sum_{w \in  T_L}\ U^{L}_w$. In other words, it aims to recruit a team with a high confidence level if recruited by leader $L$.\\
3) The third term $\sum_{w \in  T_L}C_{w,\bar{j}}$ is added to reduce the recruitment cost.\\
Finally, 4) The last term in~\eqref{scoreequation} describes the social network relationships between all workers of the team including the leader $L$. 

Note that the skill $\bar{j}$ in~\eqref{scoreequation} is set according to the combination  $s_L(\mathcal T_L)$. Also, all four terms in~\eqref{scoreequation} are normalized so they have the same order of magnitude. Consequently, $\text{TE}(L,\mathcal T_L,s_L(\mathcal T_L))$ is a dimensionless multi-objective function weighted with $\gamma_i$, $i \in \{1,\dots,4\}$. The values indicates the platform's team formation strategy.

\begin{figure}[t]
\begin{minipage}[h]{1\linewidth}
    \centering
        \hspace{-0.9cm}
    \includegraphics[width=9.5cm]{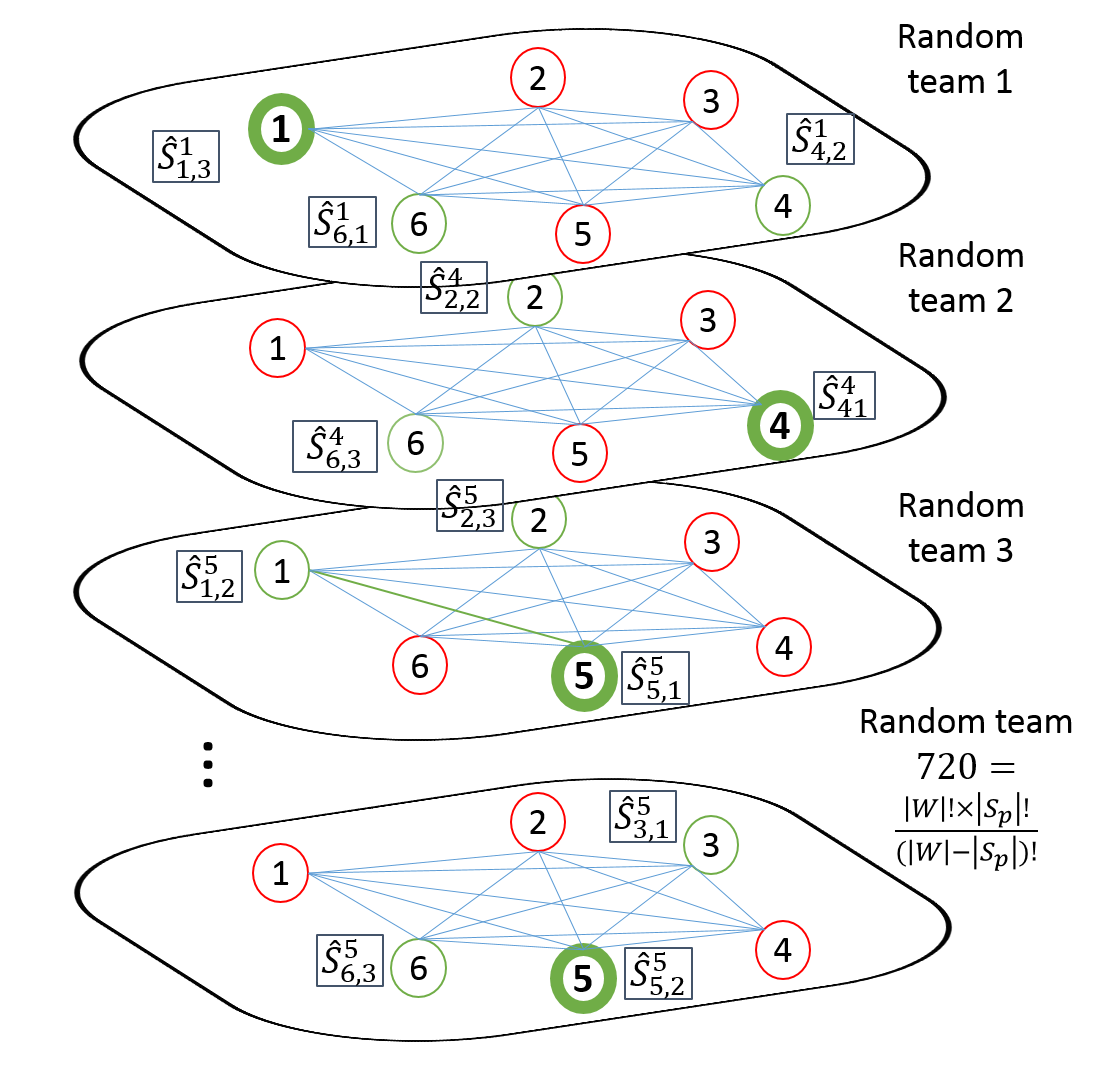}\vspace{-0.0cm}
    \caption{The process of evaluating all possible teams with $|W|=6$ workers and $|S_p|=3$ required skills.  The first team has a leader $L=1$ (bold circle), $\mathcal T_1=\{1,6,4\}$, and $s_1(\mathcal T_1)=\{S_{13},S_{61},S_{42}\}$. }
    \label{fig2}    

\end{minipage}
\end{figure}




The recruitment optimization problem is, then, written as follows:
\begin{align}
\text{(P): }  \underset{L, \mathcal T_L, s_L(T_L)}{\text{Max}}
\text{TE}(L, \mathcal T_L, s_L(T_L)).
\end{align}
This optimization problem is classified as an NP-hard problem. A brute-force technique can be used to solve it but if we enumerate all possible teams and combinations to test then, the platform, managing $|W|$ workers and looking to recruit a team for a project with $|\mathcal S_p|$ required skills, needs to go through a search area of size $\frac{|\mathcal W|!\times|\mathcal S_p|!}{(|\mathcal W| -| \mathcal S_p|)!}$. The example of shown in Fig.~\ref{fig2} with $|\mathcal W|=6$ and $|\mathcal S_p|=3$ requires 720 tests which become prohibitively large for real-world applications. Therefore, in the next section, we present a low complexity team formation stochastic approach.


\section{Stochastic Approach}\label{sec5}

The proposed probabilistic algorithm can efficiently solve the problem (P) and saves the computational resources of the cloud crowdsourcing server. The proposed algorithm uses the optimal stopping strategies and it is based on the odds-algorithm. The strategy consists of making a decision by observing multiple events one after the other and stopping on the first interesting event. In our case, the interesting event is a team that satisfies (P) with the current knowledge about the already tested teams only. 

We know exactly how many teams can be formed. We assume that all teams are equiprobable and then, we evaluate their efficiency randomly, one by one, and sort them accordingly. After evaluating any of the teams, the platform can assign it to the project but it must be careful since if an assigning decision is made there is no way to cancel it, Hence, the objective then is to select the last success team. In other words, the platforms chooses the last interesting team without verifying the remaining non-evaluated combinations.

\begin{figure}[t]
\begin{minipage}[h]{1\linewidth}
    \centering
    \vspace{0.2cm}
    \includegraphics[width=9cm]{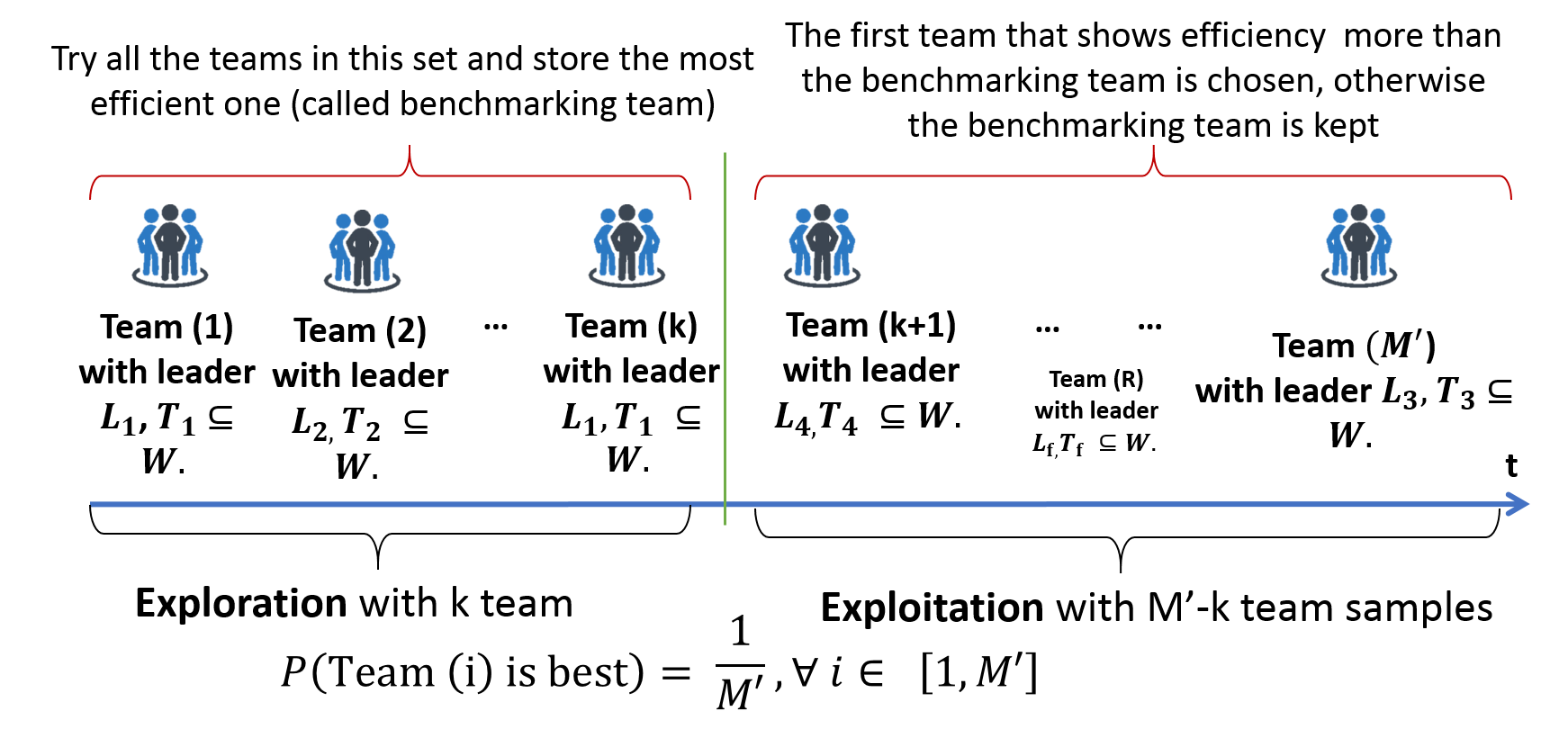}
    \caption{Stochastic algorithm for choosing a leader and a team from a set of teams with equal probability. Here, $M'=\frac{|\mathcal W|!\times|\mathcal S_p|!}{(|\mathcal W| -| \mathcal S_p|)!}$, $R \in [1,M']$, and $f \in [1,|W|]$. }
    \label{f15}
\end{minipage}
    \vspace{-0.3cm}
\end{figure}

Because teams are equiprobable, the optimum solution is to just observe the first $\frac{1}{e}\% \approx 36.8\%$ of the teams (exploration phase), then choose the first team who can achieve better efficiency than any team we encountered in the exploration phase (exploitation phase). This process is illustrated in Fig.~\ref{f15}. In the worst-case scenario, the algorithm performs all the possible combinations but the probability of going through this is $P(max  \in [1,k])=\frac{k}{n}
$ with $k=\frac{M'}{e}$ where $M'=\frac{|\mathcal W|!\times|\mathcal S_p|!}{(|\mathcal W| -| \mathcal S_p|)!}$. The value of $P(max \in [1,k])=\frac{1}{e} \approx 37\%$. Also, the probability of choosing the best team is $\approx 36.8\%$. However, the algorithm has more than $80\%$ chance to find the second best team combination.

\section{Experiments and Evaluation}\label{sec4}
In this section, we study the behavior of the proposed stochastic algorithm. We evaluate its performances using various metrics and compare them with the ones of an ILP-based optimal approach.

In order to simulate the recruitment process, we use a synthetic data with different types of projects' requirements and workers' skills. We set the values of $M=5$ and $N=14$. The uncertainty levels of potential leaders are modeled as normal distributions $\sim \mathcal{N}(0,\,0.2^2)$ and the noise levels on the skills are propositional with the number of hops between the leader and the worker. We perform Monte Carlo simulations where $1000$ realizations of different parameter settings are generated and results are averaged upon them. We also set $\gamma_i=0.25$, $\forall i$. In our experiments, all algorithms are implemented in a Python 2.6 environment and run on a 32 socket Intel(R) Xeon (R) E5-2698 v3 @2.30GHz CPU with 48G of RAM. To solve the ILP algorithm, we use the python API of academical CPLEX.
\begin{figure}
\begin{multicols}{2}
     \includegraphics[height=5.3cm]{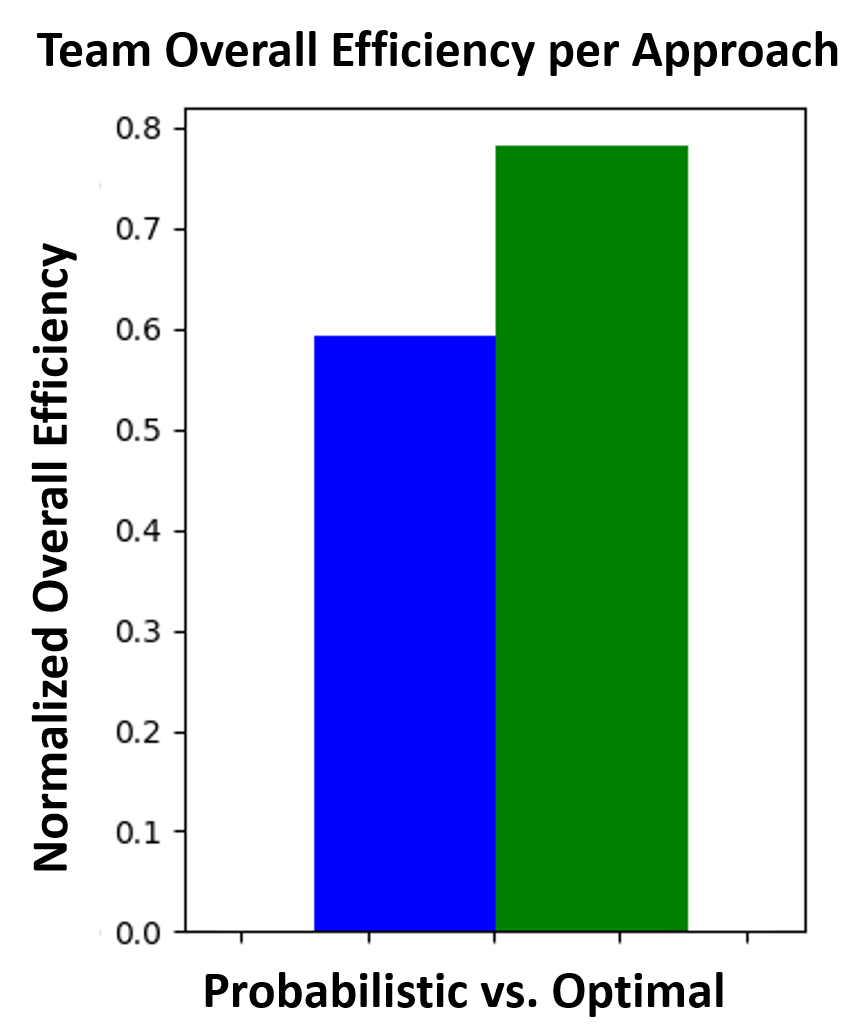}\par
         \hspace{-0.2cm}
        \includegraphics[height=5.3cm]{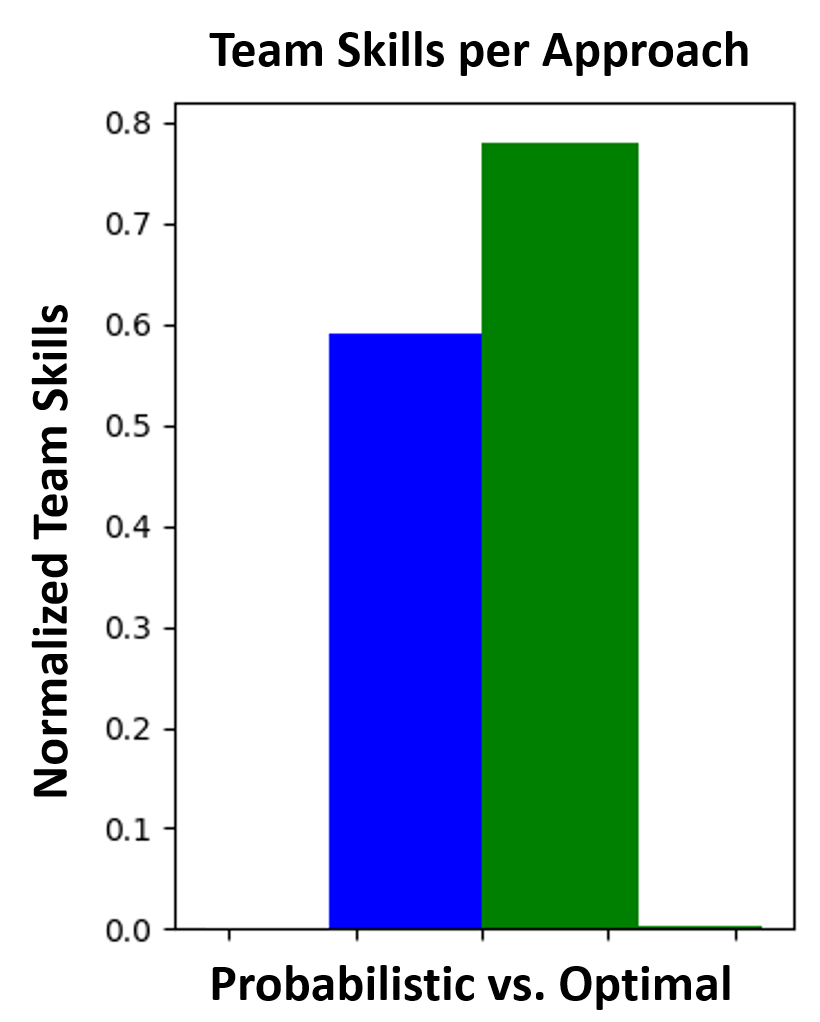}\par
    \end{multicols}
    \vspace{-1.2cm}
\begin{multicols}{2}
        \includegraphics[height=5.3cm]{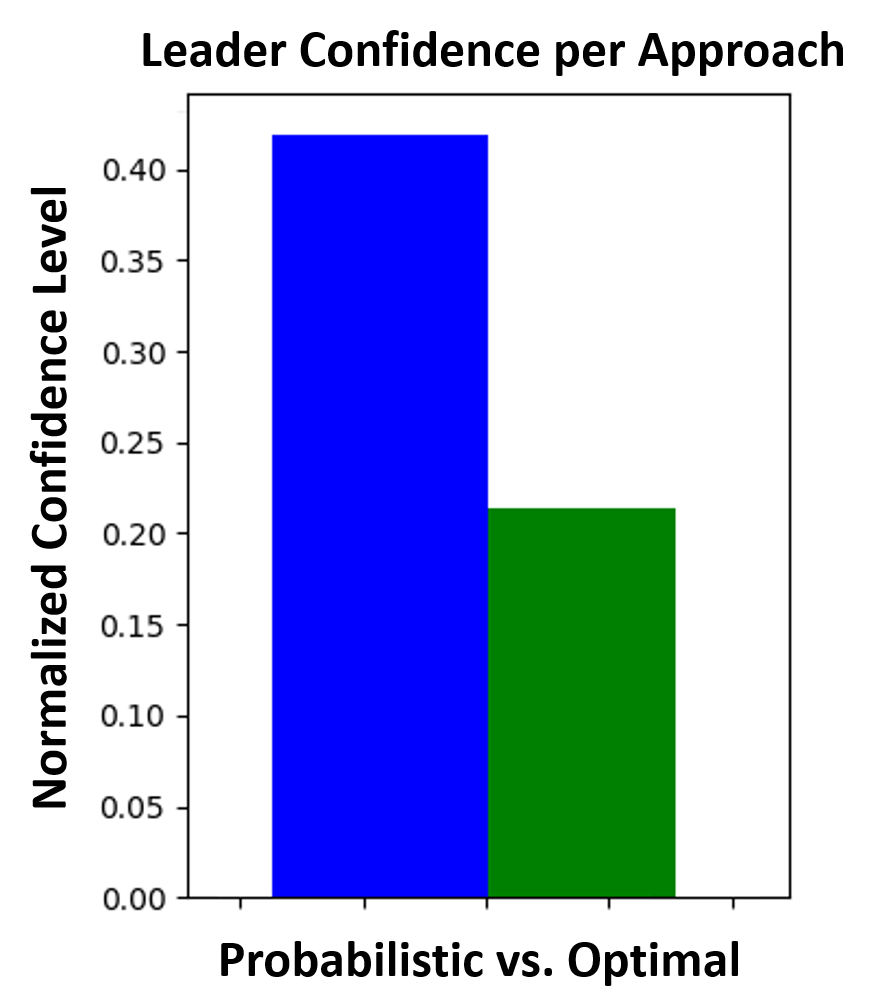}\par
                 \hspace{-0.2cm}

 \includegraphics[height=5.3cm]{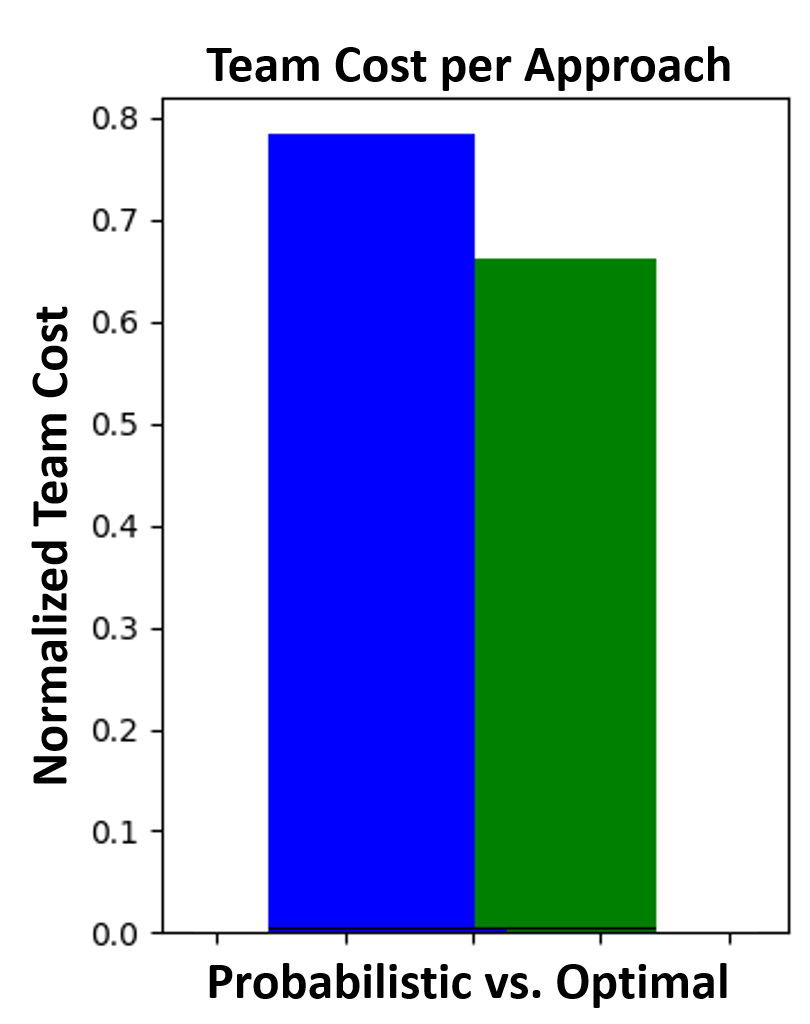}\par
\end{multicols}
    \vspace{-1.2cm}
\begin{multicols}{2}
         \includegraphics[height=5.3cm]{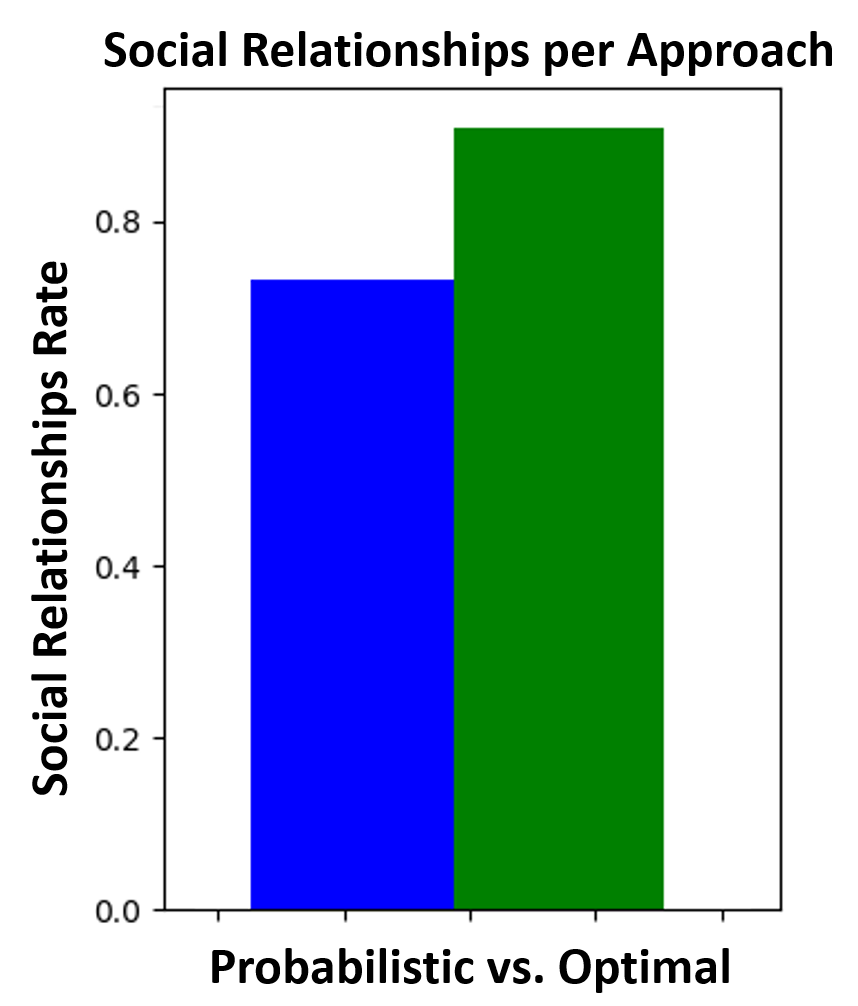}\par
                  \hspace{-0.2cm}
\includegraphics[height=5.3cm]{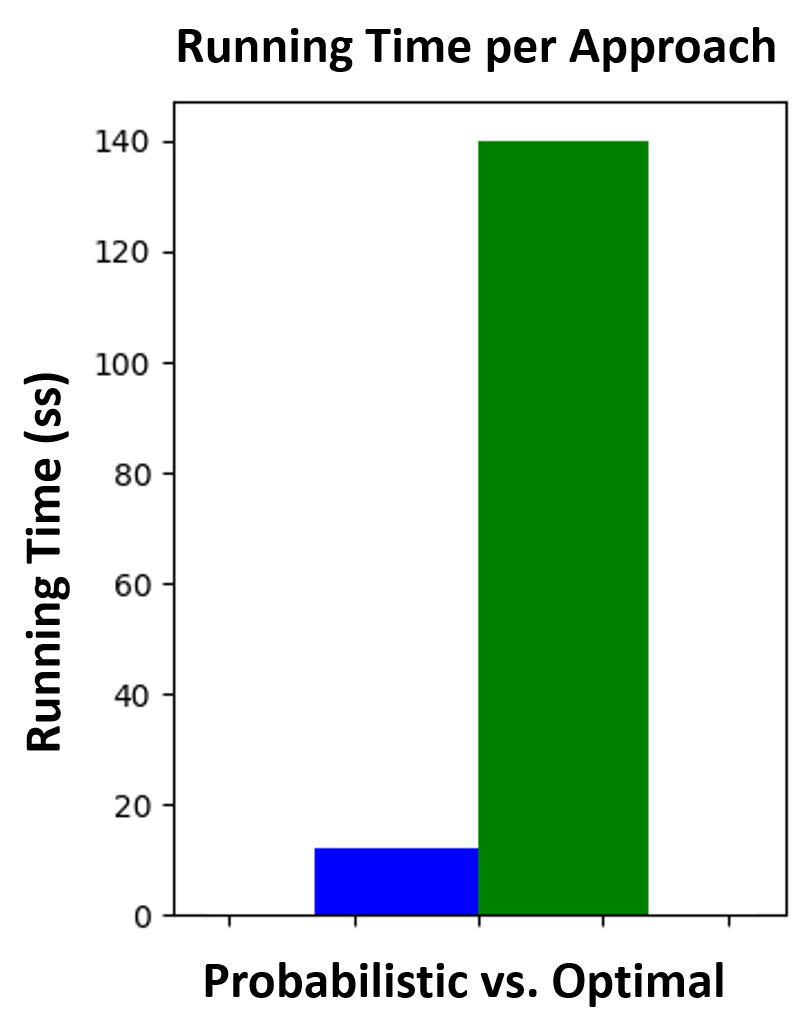}\par
\end{multicols}
    \vspace{-0.3cm}
\caption{Team overall efficiency, team skills, leader confidence, team cost, social relationship degree, and running time vs. probabilistic (left bar) and optimal approaches (right bar).}
    \vspace{-0.2cm}
\label{4figures}

\end{figure}
\begin{figure}[t]
\begin{minipage}[h]{1\linewidth}
    \centering
    \vspace{0.2cm}
    \includegraphics[width=9cm]{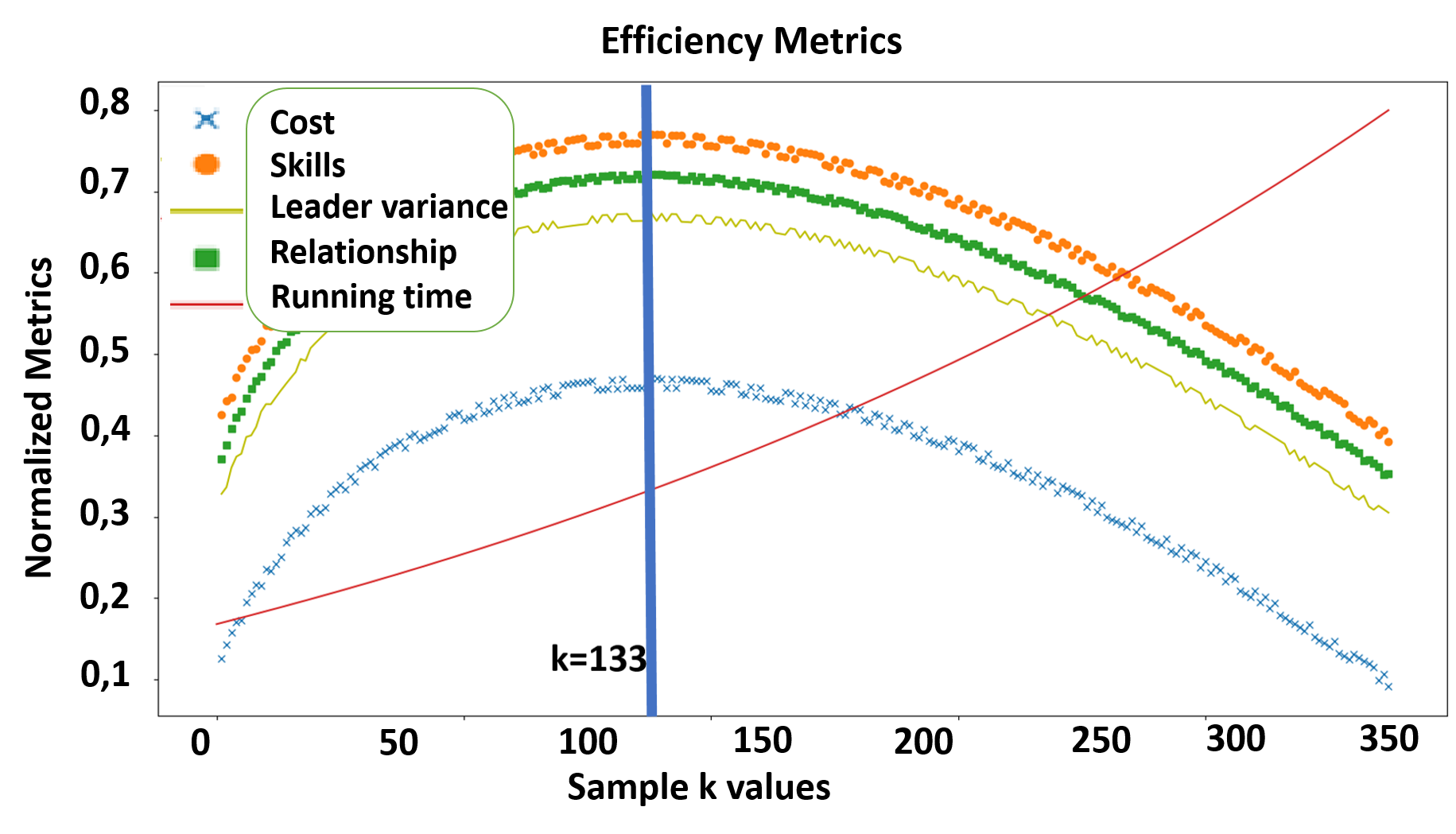}
    \caption{Efficiency metrics for a scenario where we have $|\mathcal W|=5$ and $|\mathcal S_p|=3$. The number of possible teams is $M'=\frac{|\mathcal W|!\times|\mathcal S_p|!}{(|\mathcal W| -| \mathcal S_p|)!}=360$ where we have $k=\frac{M'}{e}=133$. }
    \label{optimalstopping}
\end{minipage}
    \vspace{-0.3cm}
\end{figure}
We perform two simulations to evaluate the performance of the proposed algorithm. The first one is conducted to compare the performance of the stochastic approach against the optimal. As shown in Fig.~\ref{4figures}, we perform an average evaluation of the selected teams using the following six metrics: overall efficiency, skills efficiency, leader confidence, team cost, social relationship, and running time. The result of this simulation shows that the stochastic approach achieves close performances to ILP-based model. In fact, for example, the cost of the selected teams and the confidence levels of their leaders using the probabilistic approach is slightly higher than the optimal one. The skills efficiency of the proposed algorithm and the social relationship degree are lower than the one of the optimal solution with a gap of less than $20\%$. This is explained by the fact that although the stochastic algorithm has nearly $30\%$ chance of selecting the optimal team, it has a $\approx$ $80\%$ chance of selecting the second optimal team. 

The second simulation brings out the effect of choosing the parameter $k$ of the stochastic algorithm. The results of this simulation are illustrated in Fig.~\ref{optimalstopping} and show that for a certain values of $k$ (e.g., $k=133$ where $M'=360$), the probabilistic algorithm returns better results than other values. This corroborates the choice of $k=\frac{M'}{e}$.
\section{Conclusion}\label{sec5}
In this paper, we developed a probabilistic approach that solves team formation problems in collaborative mobile crowdsourcing frameworks using social networks.
The  proposed  algorithm  is  inspired from the optimal stopping strategies and uses the odds-algorithm to compute its output. Experimental results showed that, compared to  the  benchmark  optimal  solution,  the  proposed approach  produces reasonable performance  results with   significant computational gain.
\bibliographystyle{IEEEtran}
\bibliography{references}
\end{document}